\documentclass[preprint,showpacs,prb,endfloats]{revtex4}

\usepackage{graphicx}
\usepackage{dcolumn}
\usepackage{bm}
\usepackage{color}
\usepackage{sidecap}

\begin{document}



\title{Ultrashort dead time of photon-counting InGaAs avalanche photodiodes}

\author{A. R. Dixon}

\altaffiliation[Also at] {Cavendish Laboratory, University of Cambridge, J. J. Thomson
Avenue, Cambridge CB3 0HE, UK.}

\author {J. F. Dynes}

\author{Z. L. Yuan 
}
\email{zhiliang.yuan@crl.toshiba.co.uk}

\author {A. W. Sharpe}

\author {A. J. Bennett}

\author {A. J. Shields}

\affiliation{Toshiba Research Europe Ltd, Cambridge Research
Laboratory, 208 Cambridge Science Park, Milton Road, Cambridge, CB4~0GZ, UK }

\date{\today}

\begin{abstract}
We report a 1.036~GHz gated Geiger mode InGaAs avalanche photodiode with a detection dead time of just 1.93~ns. This is demonstrated by full recovery of the detection efficiency two gate cycles after a detection event, as well as a measured maximum detection rate of 497~MHz. As an application, we measure the second order correlation function $g^{(2)}$ of the emission from a diode laser with a single detector which works reliably at high speed owing to the extremely short dead time of the detector. The device is ideal for high bit rate fiber wavelength quantum key distribution and photonic quantum computing.
\end{abstract}


\maketitle


Many important applications, such as biological imaging,\cite{suhling02} fiber optical sensing,\cite{fibersensing} laser ranging\cite{hiskett08} and quantum communication,\cite{gobby04} require high performance single photon detectors, for which semiconductor avalanche photodiodes\cite{cova04,yoshizawa04,namekata06,yuan07} (APDs) have proved to be a practical choice due to their compactness, cryogenic-free operation and low power consumption.  Unfortunately, APDs are often associated with a long dead time, which is particularly true for InGaAs APDs. This directly limits performance of a photon counting based system. More seriously, for applications such as quantum key distribution, excessively long dead time can be exploited by an adversary, potentially compromising of the system security.\cite{makarov07,rogers07} Long dead time can be mitigated through the use of a detector array switched actively\cite{castelletto06} or passively.\cite{eraerds07} However, these solutions either add to the system complexity\cite{castelletto06} or increase the dark count rate.\cite{eraerds07}

We recently demonstrated a promising technology for high speed single photon detection with InGaAs/InP APDs.\cite{yuan07} A self-differencing (SD) circuit that compares the APD output with that in the preceding period allows detection of extremely weak avalanches, thus enabling fast, efficient, low-noise single photon detection. A 100-MHz photon count rate has been achieved, but its full potential was prevented by limitations in the photon counting electronics. Although SD-APDs have already been successfully applied to high bit rate quantum key distribution\cite{yuan08,dixon08} and quantum random number generation,\cite{dynes08} it remains desirable to experimentally determine their \textcolor{black}{ability to perform} photon counting.

In this letter, we study experimentally the photon counting dead time using a high bandwidth oscilloscope which eliminates all speed deficiencies associated with photon counting electronics. The dead time for an SD InGaAs APD operating at a gating frequency of 1.036~GHz  is determined to be 1.93~ns. Thanks to the short dead time, we demonstrate a count rate of 497~MHz at a high detection efficiency of 14\%. Such an ultra short dead time makes it possible to use a single detector in applications that generally require two or more. A single detector Hanbury-Brown and Twiss (HBT) experiment\cite{hbt} is realized to measure the second order correlation of the emission from an attenuated laser.

\begin{figure}[!b]
\centering\includegraphics[width=.88\columnwidth]{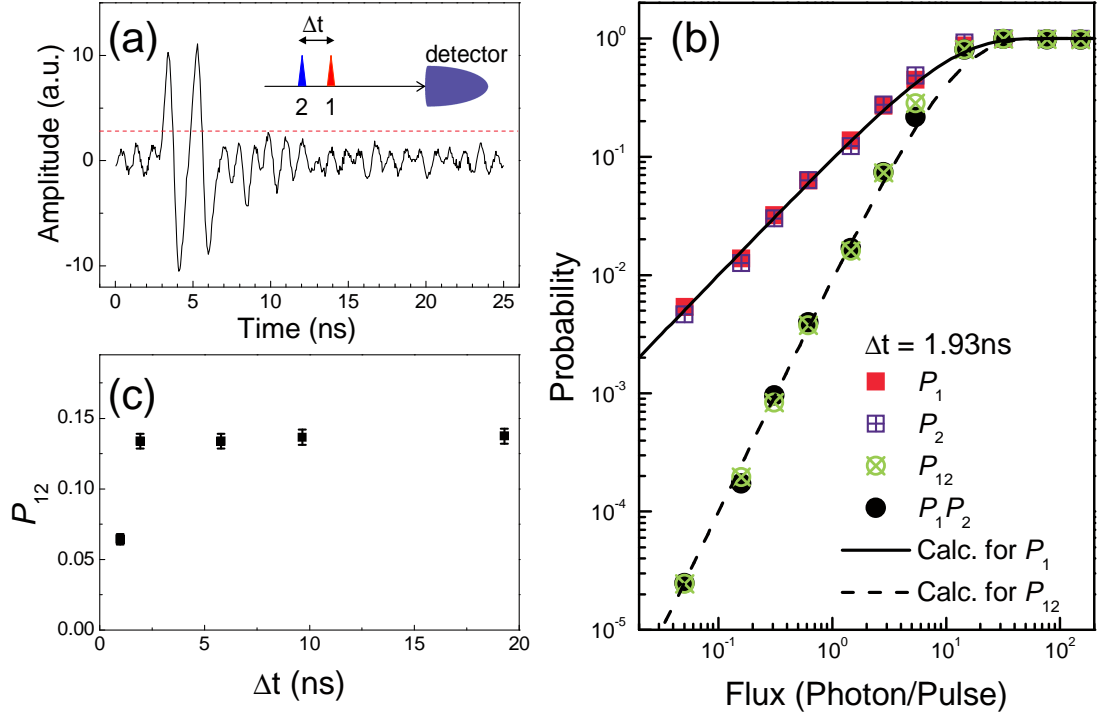}
\caption{(Color online) Experimental verification of the ultrashort detector dead time at a detector gating clock of 1.036~GHz using a double pulse method as schematically shown in the inset of (a). (a) Single shot output of an SD-APD recorded by an oscilloscope, displaying detection of both optical pulses separated by 1.93~ns. (b) Probabilities for detecting the first pulse ($P_1$), the second ($P_2$) and both simultaneously ($P_{12}$). Also shown is product of $P_1$ and $P_2$. The lines show the theoretical calculation using a detection efficiency of 10\%, assuming no influence from detector dead time. (c) Joint detection
probability as a function of the time delay between the two pulses. The illumination intensity corresponds to a single pulse detection probability of 0.36.}
\label{fig:deadtime}
\end{figure}

As an important parameter, dead time deserves a clear definition. It is usually regarded as the time after a photon detection event during which the detector is unable to register another photon. Here, we define dead time as the smallest time duration after which the detection efficiency is independent of previous photon detection history.
In other words, a detector should recover to its intrinsic detection efficiency within the dead time after a previous photon detection.
In principle, an SD-APD is able to produce an avalanche within every gating period. However, due to the self-differential nature, an avalanche signal will be canceled out if there is an avalanche in its preceding gate. This renders a theoretical dead time of two gating periods for a self-differencing detector.

To determine the dead time, we illuminate the device with a train of double pulses\cite{kerman06} with variable time separation ($\triangle t$), see Fig.~\ref{fig:deadtime}(a) inset.  The single photon detector under study is an InGaAs APD cooled electrically to $-30^\circ$C. It is operated in SD mode with a 1.036~GHz voltage square wave gating.  A 15-ps pulsed laser at a wavelength of 1550~nm is used for illumination. Two pulses of equal intensity and variable time separation, with a repetition rate of 16.19~MHz (1.036~GHz/64), are achieved through use of external intensity modulation of the pulsed laser. \textcolor{black}{Since only photons arriving within a detector gate can be detected,\cite{yuan07} the laser and the detector gating are synchronized with their relative delay tuned to ensure efficient photon detection.}
The detection efficiency is set to be around 10\% by tuning the APD DC bias. Under such conditions, the afterpulse probability is about 5\% integrated over many gates, thus becomes negligibly small for each individual gate.

We first examine the SD-APD dead time. An oscilloscope is used to record single shot waveforms of the APD output. A typical waveform displaying detection of both optical pulses is shown in Fig.~\ref{fig:deadtime}(a). Repeated measurements allow us to determine probabilities for detecting each pulse individually or both together. Detection probabilities are shown in Fig.~\ref{fig:deadtime}(b) as a function of the incident photon flux for $\triangle t=$ 1.93~ns.
Also plotted is a theoretical calculation assuming that detections for pulses 1 and 2 are independent of each other. Here we list experimental observations:
\begin{enumerate}
  \item $P_1 = P_2$ within experimental uncertainty, where $P_1$ and $P_2$ are detection probabilities for detecting pulses 1 and 2 respectively; The result $P_1 = P_2$ suggests that detector afterpulsing is negligible.
  \item $P_1$ and $P_2$ are linearly dependent on the incident photon flux before detector saturation;
  \item $P_{12}=P_1 \cdot P_2$, where $P_{12}$ is the probability for simultaneous detection of both pulses.
  \item All the measured parameters agree well with calculation using a photon detection efficiency of 10\%.
\end{enumerate}
\noindent Based on the above observations, we can deduce the conditional detection probability $P_{2|1}$, \textit{i.e.}, the detection probability of pulse 2 given pulse 1 is detected, as
\begin{equation}
P_{2|1}=P_{12}/P_1=P_1 \cdot P_2 /P_1 = P_2.
\end{equation}
\noindent The result $P_{2|1}=P_2$ states that detection of pulse 1 does not affect detection probability of pulse 2. \textcolor{black}{1.93~ns after} detection of pulse 1, the detector has already recovered to its detection efficiency of 10\%. This exactly satisfies our definition for the dead time. We can therefore conclude that the SD-APD operating at 1.036~GHz has a dead time of 1.93~ns.

Figure~\ref{fig:deadtime}(c) shows the joint detection probability $P_{12}$ as a function of the time delay $\triangle t$ measured at a fixed photon flux. The joint probability does not depend on the time delay provided that $\triangle t$ is not smaller than two gating periods, illustrating again that the experimentally determined 1.93~ns is indeed the detector dead time.  $P_{12}$ is significantly smaller when $\triangle t$ is 1 gating period, which is smaller than the detector dead time. This is due to cancelation of avalanches in consecutive gating periods by the SD circuit. $P_{12}$ does not fall to zero because of incomplete cancelation of the avalanche signals, for which the amplitude distribution is photon number dependent.\cite{kardynal08}

\begin{figure}[!t]
\centering\includegraphics[width=0.65\columnwidth]{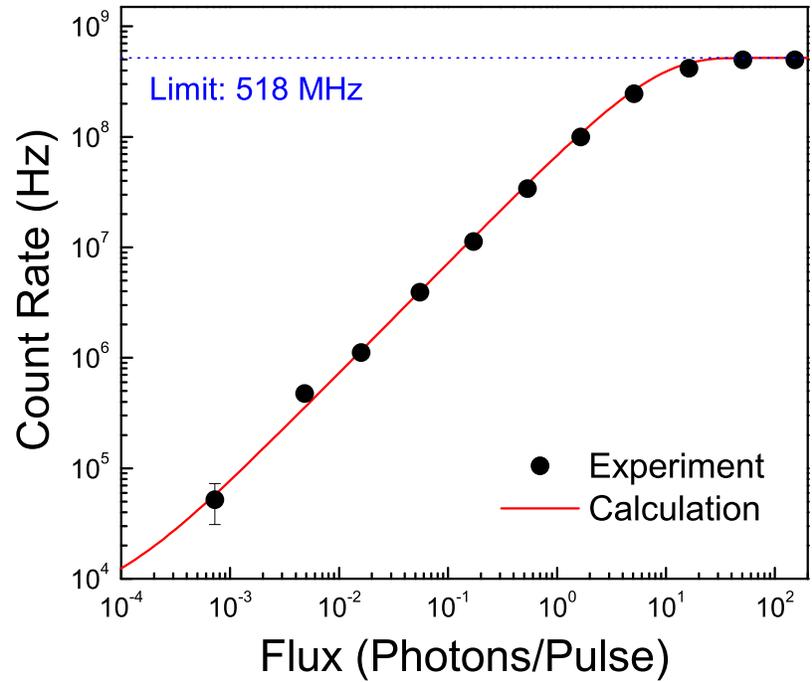}
\caption{(Color online) Experimental and calculated count rates \textit{vs.} incident photon flux. The dashed line shows the dead time limited count rate.}
\label{fig:500MHz}
\newpage
\end{figure}

A short dead time is a necessary but not sufficient condition to achieve a high count rate. To examine the detection rate limit, the SD-APD is illuminated by an attenuated 1550~nm pulsed laser with a repetition rate of 518~MHz, corresponding to an illumination pulse every other detector gate. The measured count rate is plotted in Fig.~\ref{fig:500MHz} as a function of the incident photon flux, together with a theoretical calculation using a detection efficiency of 14\% and a dark count probability of $1.67 \times 10^{-5}$ per gate. The observed count rate agrees well with the theoretical curve, both displaying a linear dependence on the incident flux before saturation. Under strong illumination ($>10$~photons/pulse), the measured count rate saturates at 497~MHz, reaching 96\% of its dead time limited capacity of 518~MHz being half the detector clock frequency of 1.036~GHz. The small discrepancy (4\%) is attributed mainly to afterpulses\cite{yuan07} in non-illuminated gates, which cancel a fraction of photon-induced avalanches.

\textcolor{black}{A few remarks are needed for the achieved dead time and count rate. (1) The demonstrated dead time is the shortest achieved so far for any semiconductor based detector system. It is more than one order of magnitude shorter than APDs operating in an active quenching mode.\cite{cova04} (2) Not only is the demonstrated 497~MHz the highest count rate achieved so far for any single element APD, but more importantly also exceeds the value obtained by an APD array of 132 \textcolor{black}{elements}.\cite{eraerds07} (3) The APD outperforms photo-multiplier tubes in photon count rate.\cite{moon08} It also compares favorably with cryogenic superconducting nanowire based detectors, whose inductance-limited recovery time usually supports a photon count rate not greater than 120~MHz at an optimized efficiency.\cite{kerman06} (4) The count rate can be further increased by operating the APD at an even higher clock frequency.
As a compact semiconductor device, APDs will play a significant role in applications where high count rate is of paramount importance.}

\begin{figure}[!t]
\centering\includegraphics[width=0.70\columnwidth]{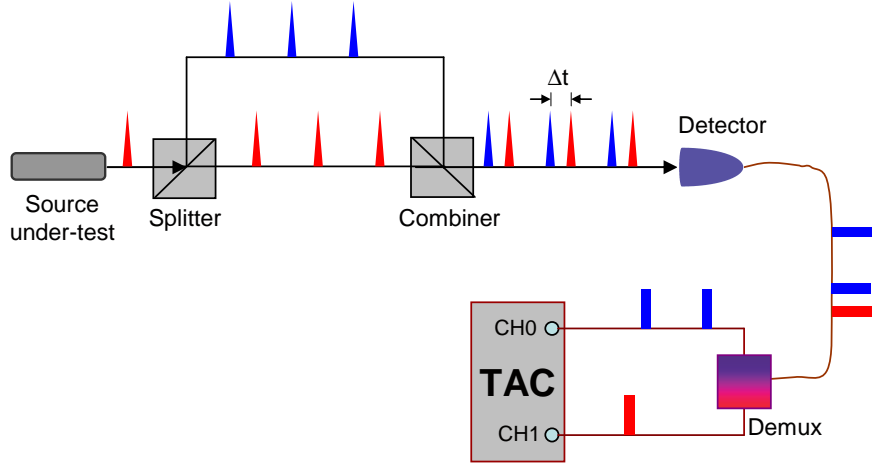}
\caption{(Color online) Configuration for the second order correlation $g^{(2)}$ measurement of a pulsed source with a single detector. Before the time acquisition card (TAC), a de-multiplexer (Demux) is used to separate electrical signals due to photons that passed through different paths of the asymmetric Mach-Zender interferometer.\\\\}
\label{fig:hbt}
\end{figure}

\begin{figure}[!t]
\centering\includegraphics[width=.70\columnwidth]{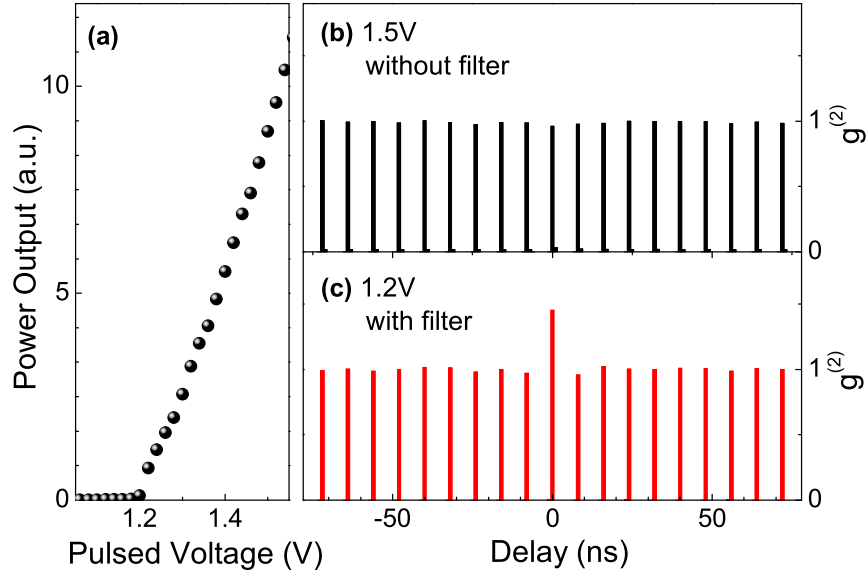}
\caption{(Color online) Second order correlation function $g^{(2)}(t)$ measurements by a single detector HBT configuration shown in Fig.~\ref{fig:hbt} for a semiconductor diode laser pulsed at 125~MHz. (a) Laser output power as a function of driving pulse voltage; (b) $g^{(2)}$ measured for the diode biased at 1.5~V; (c) $g^{(2)}$ measured for a spectrally filtered mode of the diode biased at its lasing threshold.}
\label{fig:g2}
\end{figure}

Apart from offering high bit rates,\cite{dynes08,yuan08,dixon08} a high speed device is also useful for applications that conventionally require two or more detectors.
As an illustration, a single detector is sufficient to measure the second order correlation $g^{(2)}(t)$ of a pulsed source using a modified HBT setup. The trick here is to time-multiplex optical signals so that two beams can effectively be detected by a single detector. As shown in Fig.~\ref{fig:hbt}, a pulsed source under test is split and recombined using an asymmetric Mach-Zender interferometer. The time difference $\triangle t$ between two optical arms corresponds to an integer number of detector gating periods. The recombined beam is detected by an SD-APD, whose electrical output is de-multiplexed before analysis by time correlation electronics.

A semiconductor diode laser, heavily attenuated to single photon level,  is tested in the single detector HBT setup with $\triangle t = 4$~ns.  Driven at 125~MHz by a voltage pulse generator, the laser emits a number of axial modes across a spectral region of 1540--1565~nm. Its power dependence on driving voltage indicates a lasing threshold of 1.20~V, as shown in Fig.~\ref{fig:g2}(a). When biasing the diode well above its threshold, the measured second order correlation function gives $ g^{(2)} (0) = 1.0$, as shown in Fig.~\ref{fig:g2}(b).  The unity $g^{(2)}$ is consistent with a laser source that obeys Poissonian photon number statistics for above threshold operation.

When operating the diode at its lasing threshold, \textit{i.e.}, 1.2~V, the laser intensity is expected to fluctuate. Such fluctuation can be revealed experimentally as bunching ($g^{(2)}(0)>1$).\cite{bunching} The bunching effect is more significant in the output of each individual axial mode, because competition for intensity among the modes further increases fluctuation. We therefore measure $g^{(2)}$ for a spectrally filtered mode. As expected, the measured $g^{(2)}$ displays a strong bunching at zero delay with $g^{(2)}(0)=1.4$, as shown in Fig.~\ref{fig:g2}(c).  Note that the $g^{(2)}$ measurement results are in excellent agreement with those obtained with a conventional HBT setup consisting of two detectors (data not shown here), thus corroborating that the single detector have faithfully measured the second order correlations.

In summary, we have determined experimentally an ultrashort dead time of less than 2~ns and a photon count rate of nearly 500~MHz for an InGaAs APD. The short dead time is important for high bit rate single photon applications, such as quantum key distribution and photonic quantum computing. The detector may also be used for recording second-order correlations using a single detector.

\newpage
\printfigures
\end{document}